\documentclass[iop]{emulateapj}
\usepackage{amsmath}

\def\AU{{\,\rm AU}}
\def\parsec{{\,\rm pc}}
\def\micron{{\,\rm \mu m}}
\def\erg{{\,\rm erg}}
\def\ergs{{\,\rm {erg\,s^{-1}}}}

\def\Myr{{\,\rm Myr}}
\def\Gyr{{\,\rm Gyr}}
\def\Jup{{\,\rm Jup}}

\def\hr{{\,\rm hr}}

\def\max{\rm max}
\def\diff{\rm diff}

\begin{document}
\title{Directly Imaging Tidally Powered Migrating Jupiters}
\author{Subo Dong\altaffilmark{1}, Boaz Katz\altaffilmark{2,3} 
and Aristotle Socrates\altaffilmark{2}}
\affil{Institute for Advanced Study, Princeton, NJ 08540, USA}

\altaffiltext{1}{Sagan Fellow}
\altaffiltext{2}{John Bahcall Fellow}
\altaffiltext{3}{Einstein Fellow}

\begin{abstract}
Upcoming direct imaging experiments may detect a new class of
long-period, highly luminous, tidally powered extrasolar gas giants.
Even though they are hosted by $\sim \Gyr\rm{s}$-``old' main-sequence
stars, they can be as ``hot' as young Jupiters at $\sim 100 \Myr$, the
prime targets of direct imaging surveys. They are on years-long orbits
and presently migrating to ``feed' the ``hot Jupiters'. They are
expected from ``high-$e$' migration mechanisms, in which Jupiters are
excited to highly eccentric orbits and then shrink semi-major axis by
factor of $\sim 10-100$ due to tidal dissipation at close periastron
passages. The dissipated orbital energy is converted to heat, and if
it is deposited deep enough into the atmosphere, the planet likely
radiates steadily at luminosity $L \sim 100-1000 L_{\rm Jup}  (2\times10^{-7} - 2\times10^{-6} L_{\odot})$  during a
typical $\sim \Gyr$ migration time scale. Their large orbital
separations and expected high planet-to-star flux ratios in IR make
them potentially accessible to high-contrast imaging instruments on
10m-class telescopes. $\sim 10$ such
planets are expected to exist around FGK dwarfs within $\sim
50\parsec$. Long-period RV planets are viable candidates, and the
highly eccentric planet HD 20782b at maximum angular separation $\sim
0.08\arcsec$ is a promising candidate. Directly imaging these
tidally powered Jupiters would enable a direct test of high-$e$
migration mechanisms. Once detected, the luminosity would provide a
direct measurement of the migration rate, and together with mass (and
possibly radius) estimate, they would serve as a laboratory to study
planetary spectral formation and tidal physics.
\end{abstract}

\section{Introduction}
\label{sec:intro}

Jupiter and Saturn analogs orbiting other Sun-like main-sequence 
stars have evaded direct detection. With an effective 
temperature of $\sim 124 {\rm K}$, an extrasolar analogue of Jupiter is 
fainter than a Sun-like host by $\gtrsim 10^{8}$ in 
near-IR \citep{kasting}, beyond the reach of instrument capabilities at 
present and in the near future. While ``hot Jupiters'' 
(Jovian planets at $\lesssim 0.1 {\AU}$) have high temperatures and 
thus large planet/star flux ratios, they are too close to their hosts 
($\ll0.1''$) to be spatially resolved by $10$m-class telescopes. 
To date, direct-imaging surveys have focused on searching for long-period 
massive gas giants around nearby young stars with ages 
$\lesssim 100 \Myr$, a strategy that has led to a number of discoveries 
(e.g., \citealt{hr8799,lag10}).
In these systems, the planets are still ``hot'' -- cooling down 
from presumably high temperatures at birth, which significantly 
enhances their flux ratios with host stars.

In this Letter, we discuss the possibility of directly imaging 
a third class of  ``hot'' gas giants  (besides close-in hot 
Jupiters and young Jupiters around young stars), consisting of 
a population of long-period, very luminous, tidally-powered  
planets undergoing orbital migration.

\section{High-eccentricity Migration due to Tidal Dissipation}

It is commonly believed that progenitors of hot Jupiters are formed 
with semi-major axis $a$ beyond the ``snow line'' at a few $\AU$ 
and then migrate inward to their current locations by shrinking 
$a$ by factor of $\sim 10 - 100$ \citep{lin}. One class of proposed migration 
mechanisms involve exciting long-period Jupiters to highly 
eccentric orbits, due to gravitational interactions with stellar or 
planetary perturbers, enabling them to lose orbital energy at 
successive close periastron passages through tidal interactions 
with their host stars. Such ``high-$e$'' mechanisms include 
Kozai-Lidov Cycles plus Tidal Friction (KCTF) 
(\citealt{wu03,fab07}), 
planet-planet scatter \citep{ras96}, ``secular chaos'' \citep{wu11} 
etc. Recently high-$e$ mechanisms have gained 
observational support. A significant fraction of hot Jupiters 
are found to be on misaligned orbits with respect to their host stars' 
spin axes \citep{win10,tri10}, which is a natural consequence of such 
mechanisms \citep{fab07}. 

One general expectation from all high-$e$ mechanisms is that there 
should exist a steady-state migrating population of long-period, 
highly eccentric gas giants ``feeding'' the hot Jupiters 
\citep{soc12}. This results from the continuous 
generation of hot-Jupiter progenitors due to constant  
formation of stars and their planetary systems over the age of 
the Galaxy.
The orbital angular momentum is approximately conserved 
during tidal dissipation, so the actively migrating Jupiters 
have $a(1-e^2) \equiv a_{\rm F}$, where $a_{\rm F} \la 0.1 \AU$ is 
the semi-major axis of their final circularized orbit. 
According to \citet{soc12}, the frequency of this population is 
likely an increasing function of their period (and the eccentricity), 
extending to that of their ``source'' (possibly at $\gtrsim 5 \AU$). 
A possible archetype of the migrating population is HD 80606b 
\citep{hd806061, hd806062}, which is a $4 M_\Jup$
planet at semi-major axis $a=0.45\AU$ and $e=0.93$ ($a_{\rm F}=0.06 \AU$),
accompanied by an solar-mass companion at $\sim 1200 \AU$.

Regardless of the specifics of the high-$e$ mechanisms, 
a gas giant that migrates from semi-major axis $a'$ 
to $a$ over a time $\Delta{t}_{\rm m}$ loses orbital 
energy due to tidal dissipation, which is converted to heat and radiated 
away. This leads to an averaged tidal luminosity,
\begin{eqnarray}
L_m &=& \frac{GM_*M_p}{2\Delta{t}_m}\big({\frac{1}{a} - \frac{1}{a'}}\big)\cr
    &\sim& 8\times10^{26} \ergs
\big(\frac{M_*}{M_\odot}\big)
\big(\frac{M_p}{3 M_{\Jup}}\big)
\big(\frac{\Delta{a_m^{-1}}}{1 \rm AU^{-1}}\big)
\big(\frac{\Delta{t}_{\rm m}}{1 \Gyr}\big)^{-1},
\end{eqnarray}
where $M_*$ is the mass of the host star, $M_p$ is 
the mass of the planet, $\Delta{a_m^{-1}} = 1/a - 1/a'$. For
comparison, $L_\Jup = 8.6\times 10^{24}\erg\,s^{-1}$  is the 
luminosity of our Jupiter. 
\citet{fab07} presents a possible migration
path for HD 80606b due to Kozai-Lidov Cycles plus Tidal Friction (KCTF), 
and in their simulation,
over $\sim 0.1 \Gyr$, the ``migration rate'' 
$\Delta{a_m^{-1}}/\Delta{t}_{\rm m}$ at $0.5\AU$, $1\AU$, and $2\AU$ 
is $\sim 12$, $5$, and $1.4 \AU^{-1}\Gyr^{-1}$,
{\rm corresponding to $L_m \sim 1.3\times 10^{28}, 5.3\times 10^{27}, 
1.5\times10^{27}\ergs$, respectively, 
which span $\sim 1500 - 180 \,L_{\Jup}$ (see also \citealt{wu03}).} If tidal dissipation 
occurs in a deep enough layer of the planet atmosphere, 
the thermal relaxation time $t_{\rm th}$ can be much longer than the 
orbital time scale ($P \sim \rm yr$), the planet would constantly
radiate at its tidal luminosity. The physics of tidal dissipation in giant planets
is an unsolved theoretical problem, and there is no reliable method to 
calculate where the tidal dissipation happens in the planet atmosphere. 
The upper limit of $t_{\rm th}$ is the Kevin-Helmholtz time scale 
$t_{\rm KH}  \propto M_p^2/R_p/L_p$,
which is about $0.1 \Gyr$  for a Jupiter with luminosity $L_p$ at hundreds of $L_{\Jup}$.
For the Jovian planets of interest, $M_p^2/R_p$ is at most factor of 
100 smaller than that for the Jupiter, and at this extreme, for 
$L_p = 100\,L_{\Jup}$, $T_{\rm KH}$ is still $\sim {\rm Myr}$, much longer than the orbital time scale. 
The tidal forcing acts on the planet body as a whole, so it is reasonable to expect that significant tidal dissipation 
operates at a depth that is a considerable fraction of the planet 
radius and then $t_{\rm th}$ is approximately $t_{\rm KH}$, which is many 
orders of magnitude larger than the orbital time. 
We make the reasonable
assumption $t_{\rm th} > P$ throughout the paper.

The luminosities of migrating Jupiters can be tidally enhanced by 
$\sim 2-3$ orders of magnitude, comparable to  
young Jupiters at $\sim 100 \Myr$ \citep{bur97, bar03, mar07, spiegel}, 
which are the prime targets for ongoing and planned direct-imaging surveys 
\citep{gemini, nici}. These tidally powered Jupiters could be located
at $a \sim {\rm several} \AU$, and their maximum separations at 
apastron are further enhanced by high eccentricity $e \sim 1$ by 
a factor $(1+e)\simeq 2$, making them promising targets for 
direct-imaging detections. 

Tidally powered Jupiters are not necessarily limited to those 
actively migrating at high eccentricity. For example, in the 
specific cases of KCTF, while the planet visits the highly 
eccentric phase that enables tidal dissipation at periastron 
passages in each Kozai-Lidov cycle, it typically spends much 
longer (factor of $\sim 10$ more, see \citet{fab07}) time oscillating 
at low-eccentricity orbits. The oscillation amplitude in $e$
is generally larger when the planets are at longer period where
relativistic precession is weaker. If the thermal time 
scale $t_{\rm th}$ is longer than the Kozai-Lidov time 
scale ($t_{\rm Kozai} \sim 0.02 \Gyr$ for HD 80606b while 
migrating at $\sim 5\AU$), the planet radiates approximately at 
the averaged tidal luminosity $L_m$ during the whole Kozai-Lidov 
cycle. Therefore, the tidally-powered Jupiters include not only 
those at high $e$ (i.e., small pericenter) but also a factor of $\sim 10$ 
more experiencing Kozai-Lidov oscillations at lower $e$.

\begin{figure}[h]
\epsscale{1.2} \plotone{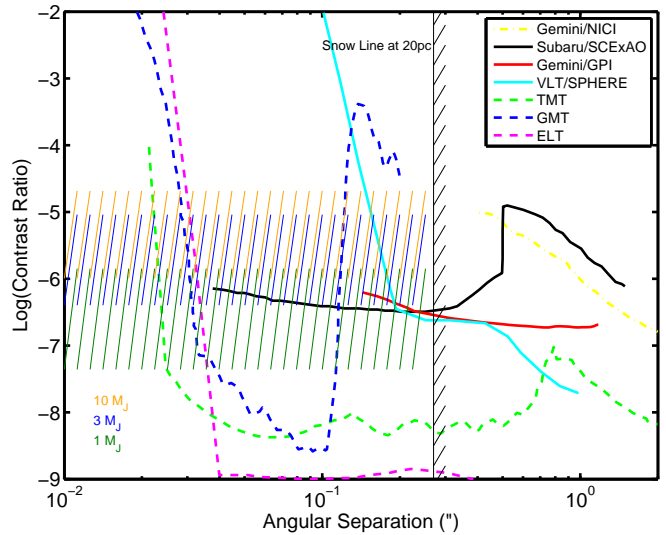}
\caption{The planet-to-star contrast ratios for expected 
tidally-powered Jupiter population as a function of angular separations 
as compared to those reached or expected for current (dotted-dash lines) and upcoming (solid lines) high-contrast imaging instruments 
for 10m-class telescopes as well as future instruments for 30m-class telescopes (dash lines). The expected $H$-band contrast ratios at a range of migration rate of $1-10 \AU^{-1}\Gyr^{-1}$ for 
$1, 3$ and $10$ $M_J$ planets are shown in dark green, blue 
and yellow shaded regions, respectively. They are calculated using 
the models in \citet{bur97} as discussed in the text. 
The maximum angular separation (at apocenter) for a 
super-eccentric with $a = 2.7\AU$ (snow line for a solar-mass host)  
at 20pc is shown as vertical black line with shading, indicating 
a plausible upper boundary in orbital separations for these Jupiters. Upcoming instrument such as Subaru/SCExAO may have substantial sensitivity at small enough angular separations to probe the tidally powered Jupiter population, and high-contrast imaging instruments on future $30$-m class telescopes (such as TMT, GMT and ELT) are expected to have excellent capabilities to detect such planets.
The sources for the contrast curves are listed below.  (1) Gemini/NICI $H$-band: \citet{nici} 
(2) Subaru/SCExAO with extreme AO, $1-\hr$ integration: http://www.naoj.org/Projects/newdev/intmtg/20100426/files/SCExAO\_overview\_2010-04-26.pdf (3) Gemini/GPI -- http://planetimager.org/pages/gpi\_tech\_contrast.html (4) VLT SPHERE and ELT EPICS in $J$-band: \citet{eelt} (5) TMT Planet Formation Instrument at $1.65 \micron$: https://e-reports-ext.llnl.gov/pdf/333450.pdf (6) GMT HRCAM at $1.6 \micron$ with
$>1 \hr$ exposure: http://www.physics.berkeley.edu/research/genzel/Physics250.5/literature-talks/ELT/GMT\_Science\_Case.pdf
 }
\end{figure}

\section{Direct Imaging Observations}
\label{sec:direct}
The achievable sensitivity of high-contrast imaging instruments 
degrades substantially within the so-called inner working angle, 
which is often quoted to be $\sim 2-4$ times the diffraction limit, 
$\theta_{\diff} \sim \lambda/D \sim 0.02\arcsec 
(\lambda/1 \micron)(D/10 {\rm m})^{-1}$, where $\lambda$ is the observed 
wavelength and $D$ is telescope aperture. Several high-contrast 
imaging instruments are and will be commissioned in the near future 
on a number of 8-10m telescopes, such as Gemini GPI \citep{gpi}, 
VLT SPHERE \citep{sphere}, and Subaru SCExAO \citep{seeds}, 
LBTI \citep{hin08}. The best contrast goals of these instruments 
are $10^{-7} - 10^{-8}$ in near-IR beyond the inner working 
angle \citep{gpi, sphere}. See Fig. 1 for their expected contrast ratio curves in near IR.

At a migration rate of $\Delta{a_m^{-1}}/\Delta{t}_{\rm m} = 1 - 10 
\AU^{-1}\Gyr^{-1}$, tidal luminosities for $\sim 3 M_{\Jup}$ planets 
are $\sim 2\times10^{-7} - 2\times10^{-6} L_\odot$ ($\sim 10^2$ - $10^3 L_{\Jup}$), corresponding to blackbody effective temperatures 
$T_{\rm eff} \sim 390K - 690K$ for a planet radius
$1 R_{Jup}$. The peaks of the black body radiation at these temperatures 
are at $\sim 9-5 \micron$.
The black body contrast ratios at $3.78 \micron$ ($L'$-band) 
of the planet to a Sun-like star ($T_{\rm eff} \sim 5800 K$) are 
$\sim 5.0\times 10^{-7} - 3.8\times 10^{-5}$. 
At bands further away from the peak (e.g., $\sim 1 \micron$), 
black body radiation is exponentially suppressed,  
resulting in very low flux ratios. However, the near-IR spectral energy
distribution is unlikely to be well described by black body emission. 
For example, Jupiter has a spectral window allowing for
probing deeper, warmer layers of the atmosphere 
that leads to orders of magnitude larger flux than that of 
a $125 \rm K$ blackbody in near-IR \citep{kasting}. 
We note that according to \citet{bur97}'s model, at effective 
temperatures $400K$ and $600K$, with surface gravity 
$10^4 {\rm cm\,s^{-2}}$, the contrast ratios in [$J$, $H$, $K$, $L'$]-band 
with a Sun-like star 
are approximately $[3.6\times10^{-7}, 4\times10^{-7}, 3.9\times10^{-8}, 4\times10^{-6}]$,
$[3.5\times10^{-6}, 4.8\times10^{-6}, 1.4\times10^{-6}, 2.2\times 10^{-5}]$, respectively.
Other models \citep{bar03, spiegel} generally predict similar contrast ratios in 
$L'$-band, but the predictions can vary by order of magnitude 
in $J$-band and $H$-band, possibly reflecting theoretical uncertainties 
in such calculations such as the treatment of clouds and planet luminosity evolution.

\citet{soc12} estimate that the frequency of 
long-period (hundreds of days), highly-eccentric 
Jupiters is about $10\%$ that of hot Jupiters, whose 
occurrence rate is estimated to be $\sim 1\%$ around FGK dwarfs 
\citep{marcy}. Therefore, $\sim 0.1\%$ of solar-type stars may host this 
population. There are $\sim 10^4$ FGK dwarfs within $\sim 50 \parsec$
of the Sun, which amounts to $\sim 10$ potentially tidally-powered
Jupiters. The closest super-eccentric Jupiter host is then at $\sim 20\parsec$, and suppose it starts migration near the snow line at $\sim 2.7\AU$, its 
maximum angular separation is $\sim 0.27\arcsec$ (achieved at 
apastron for $e \sim 1$), which is marked as a shaded black line in Fig.1.

In Fig.1, the expected ranges of $H$-band contrast ratio assuming 
migration rate varying from $1 \AU^{-1}\Gyr^{-1}$ to $10 \AU^{-1}\Gyr^{-1}$
for planets with $1, 3, 10 M_J$ are shown in shaded regions in green, blue and
yellow, respectively. It is challenging for present and most of the upcoming 
high-contrast imaging instruments such as Gemini/NICI, Gemini/GPI and 
VLT/SPHERE to detect the population of tidally-powered Jupiters due to their
relatively large inner working angles ($\gtrsim 0.2 - 0.3 \arcsec$). The most promising instrument is SCExAO at Subaru. The extreme AO system of 
SCExAO is expected to allow for significant sensitivity inside $0.1\arcsec$, 
which may potentially probe the tidal Jupiters as close as $a \sim 0.5 \AU$
at $20 \parsec$.

One strategy to identify these 
tidally-powered Jupiters is to follow up long-period, 
highly eccentric planets with known radial velocity (RV) orbits,  
which are likely to be actively migrating.
For $e \sim 1$, one obtains maximum projected separation very close 
to apastron at $r_{\perp, \max} \approx 2a 
\sqrt{\sin^2\omega \cos^2 i + \cos^2\omega}$, where $\omega$ is 
the argument of periastron. There is one RV planet, with 
$a\sim 1\AU$ and $a_{\rm F}\lesssim 0.1\AU$, known as   
HD 20782b \citep{jon06,too09} at $36 \parsec$, and the best-fit 
parameters are $M \sin i=1.9 M_{\Jup}$, $a = 1.38\AU$, $e = 0.97$, 
$\omega = 148^o$ (note that the eccentricity needs to confirmed as 
the periastron passage was not sufficiently probed).
It reaches maximum angular 
separation at $(0.28\cos^2 i + 0.72) 0.076\arcsec$ at apocenter, 
which corresponds to $\lesssim 2.9 \lambda/D$ at $J$-band, 
$\lesssim 2.2 \lambda/D$ at $H$-band, $\lesssim 1.7 \lambda/D$ 
at $K$-band and $\sim\lambda/D$ at $L'$-band for a 10m telescope.
The target is difficult to image for most of the current and 
upcoming high-contrast imaging instruments but it is likely to be 
accessible by SCExAO at Subaru given its small expected inner 
working angle allowed by its extreme AO system (see Fig.1). 
The long baseline of LBTI (22.8m) may be advantageous in terms of spatial
resolution, but it is challenging given the inner working angle achievable for 
the current available instrument. It may be accessible when the LBTI 
high contrast imaging system is refined (Private Communication, 
Philip Hinz 2012).

As previously discussed, if KCTF takes place, 
many long-period, low-$e$ Jupiters can be tidally-powered 
 by radiating heat accumulated from past high-$e$ visits, and
 their occurrence rate can be $\sim 10$ larger than that of the 
 actively-migrating, high-$e$ population.
A possible observation strategy would be to directly image all 
RV planets with sufficiently large maximum 
angular separations and probe this population 
(as well as those that show a linear trend, which indicates 
the probable presence of long-period planets). Planetary systems with 
known perturbers from RV residuals or imaging may receive 
preference since the planets in these system have a higher 
probability of Kozai-Lidov oscillation. Advanced AO/coronagraphs 
and better post processing techniques (e.g., \citealt{loci}) may 
help to increase the contrast and decrease the inner working angle.

It may also be possible 
to remove the speckle noise more efficiently and improve the detection
sensitivity with information from the known orbital phase from RV,
 which will further boost the detectable contrast ratio and obtain
 smaller inner working angle. 
 
A large fraction of nearby main-sequence stars have not 
been monitored by RV sufficiently long to find long-period 
Jupiters. The all-sky astrometric mission Gaia is 
expected to be sensitive to planets more massive than Jupiter 
between $\sim 0.5 - 3 \AU$ for all solar-type hosts within $50 \parsec$ 
\citep{gaia}, so it will make a nearly complete census of tidally-powered 
Jupiters in the solar neighborhood and provide an excellent sample for 
direct-imaging follow-ups.

\section{Discussion and Future Prospect}
If the tidally-powered Jupiters are directly imaged, their 
luminosities provide a direct measure of planet migration rate 
due to tidal dissipation and thus constrain high-$e$ migration 
mechanisms.\footnote{Note however thermal tidal power generated
at pericenter passages could be responsible for tidally powering
long-period Jupiters \citep{soc09a,soc09b,soc10}. In that 
case, the source of energy is from star light rather than orbit.}
Combined with RV orbits, one can break the inclination 
degeneracy in RV to obtain the de-projected true mass and full 
orbital solution, making them an excellent laboratory to study 
planetary dynamics. With measured mass and luminosity, one may
probe the spectral formation of gas giant atmosphere as well as 
the physics of tidal dissipation. 
It is interesting to note that, even though similar 
high-$e$ population may exist for binary stars  
\citep{dong12}, it is much more difficult to measure tidal luminosity 
directly due to nuclear burning.

As discussed in \citet{soc12} and \citet{joh12}, transit surveys such as 
{\it Kepler} are ideal to find the eccentric migrating Jupiters 
due to their enhanced transit probability. 
Space-based transit surveys that target bright stars 
may potentially provide an excellent sample of tidally-powered 
planets hosted by nearby stars that are suitable for direct-imaging 
study. For transiting planets, 
high-precision IR light curves with secondary eclipse could 
in principle directly measure the tidal luminosity for these 
planets as well (see \citealt{lau09}). 

Ground-based high-contrast imaging instruments are 
experiencing rapid development over the last few years, and 
the tidally powered Jupiters may turn out to be the most
 luminous planets to image  for nearby solar-type main-sequence stars. 
New instruments such as SCExAO should already be able to probe 
this population. Future telescopes such as TMT, GMT and ELT can 
conduct a thorough survey for this population 
around nearby stars due to their smaller diffraction limits (see Fig. 1). 
\acknowledgements
We thank Andy Gould, Cullen Blake, Jose Prieto, 
Scott Tremaine, Matias Zaldarriaga, Dave Spiegel, Rashid Sunyaev, 
Sasha Hinkley, Adam Burrows, and Phil Hinz for discussions. 
Work by SD was performed under 
contract with the California Institute of Technology (Caltech) 
funded by NASA through the Sagan Fellowship Program. BK is supported 
by NASA through the Einstein Postdoctoral Fellowship awarded by Chandra 
X-ray Center, which is operated by the Smithsonian Astrophysical 
Observatory for NASA under contract NAS8-03060. AS acknowledges
support from a John N. Bahcall Fellowship awarded at the 
Institute for Advanced Study, Princeton.

\end{document}